\documentclass{jpsj2}
\title{%
Scaling Property of the F-AF Spin Chain Near 
the Exactly Solvable Point
}

\author{%
Hidenori \textsc{Suzuki}\thanks{Present address: Department of Physics, College of Humanities and Sciences, Nihon University, Setagaya-ku, Tokyo 156-8550, JAPAN}  
and Ken'ichi \textsc{Takano}
}

\inst{%
Toyota Technological Institute,
Tenpaku-ku, Nagoya, 468-8511
}

\recdate{\today}

\abst{%
We investigate the ground state of the $J_1$-$J_2$ spin-1/2 chain 
with $J_1<0$ and $J_2>0$ 
in the case that the nearest-neighbor $J_1$ interaction 
in the $z$-direction has a weak anisotropy as $J_1 (1-\alpha)$. 
We perform a perturbational analysis for small $\alpha$ and 
$\lambda \equiv J_2- |J_1|/4$ with the exact solution of 
the unperturbed ground state for $\alpha = \lambda = 0$. 
The scaling property of the ground state energy is examined in detail.
By the numerical diagonalization analysis of finite size systems, 
we found the phase boundary 
equation between the spin fluid and dimer phases 
as $\alpha_{\rm c} = 14 \lambda_{\rm c}^{4/3}$. 
}

\kword{%
quantum spin chain, frustration, exact ground state, scaling property
}

\begin{document}	
\maketitle 		

\section{Introduction}
Low-dimensional quantum spin systems have attracted great attention for many years.
Among them, quantum spin chains with competing interactions have been fascinating subjects. 
This is mainly because these systems exhibit rich varieties of exotic ground states and 
phenomena owing to the quantum fluctuation enhanced by geometrical frustrations. 
For example, spontaneous dimerization\cite{majumder69},
quantum chiral phases \cite{nersesyan98, kaburagi99, allen00, aligia00, 
hikihara00, kolezhuk00, nishiyama00, hikihara01, kolezhuk02, furukawa08}, 
1/3-plateau \cite{okunishi03a, okunishi03b, tonegawa04, hida05}, 
and singlet cluster solid \cite{takano07,hida08} have been reported. 

In theoretical studies of quantum spin systems, exact solutions are 
useful for constructing physical pictures of the systems. 
However, it is generally difficult to find an exact ground state of a frustrated quantum spin system.
Despite this, exact ground states have been found in several models. 
For example, the exact dimer ground state for the antiferromagnetic $J_1$-$J_2$ chain\cite{majumder69} is well known. 
Other examples are exact spin-cluster ground states for the pure and mixed diamond chains\cite{takano96,takano08,hida09}. 
Even in the cases that the exact ground states are found, there still remains the difficulty to calculate correlation functions using them.

\begin{figure}[tbp]
\begin{center}
\includegraphics[width=80mm ]{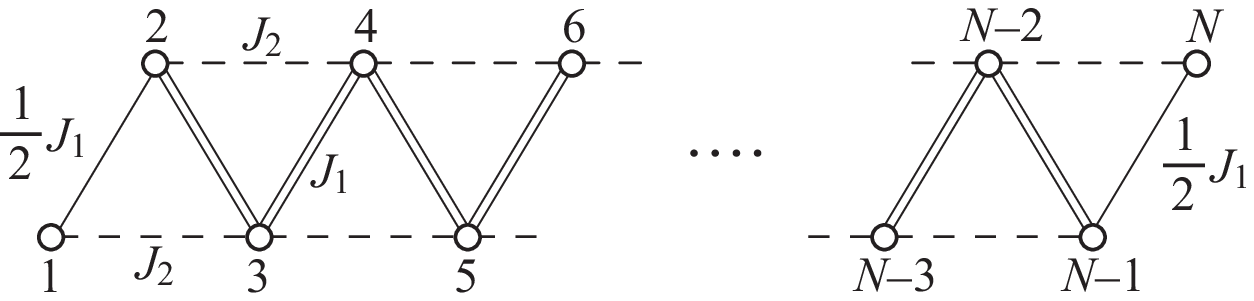}
\caption{F-AF chain with the special open boundary condition.}
\label{fig:chain}
\end{center}
\end{figure}

Here, we are concerned with the F-AF chain depicted in Fig.~\ref{fig:chain} as another example that the exact ground state is known. 
This chain is a $J_1$-$J_2$ spin-1/2 chain with ferromagnetic (F) nearest-neighbor (NN) and antiferromagnetic (AF) next-nearest-neighbor (NNN) interactions ($J_1<0$ and $J_2>0$).
The ground state is fully ferromagnetic for $J_2<|J_1|/4$\cite{niemeijer71,ono72,bader79}, while it is a singlet state for $J_2>|J_1|/4$.
At the point of $J_2=|J_1|/4$, the explicit formulae of 
ground states have been found and studied\cite{hamada88,dmitriev97a,dmitriev97b,hs08}. 
In particular, Hamada {\it et al.} first found an exact solution in a resonating valence bond form\cite{hamada88}.
Recently, we have reported all the degenerate ground states in the explicit formulae and further proposed the recursion formulae of the exact solutions.\cite{hs08}
These exact solutions are extended to include the case with bond alternations in the NN and NNN interactions. 
The greatest merit of the recursion formulae is  
that physical quantities such as correlation functions can be calculated using them. 

\begin{figure}[tbp]
\begin{center}
\includegraphics[width=65mm]{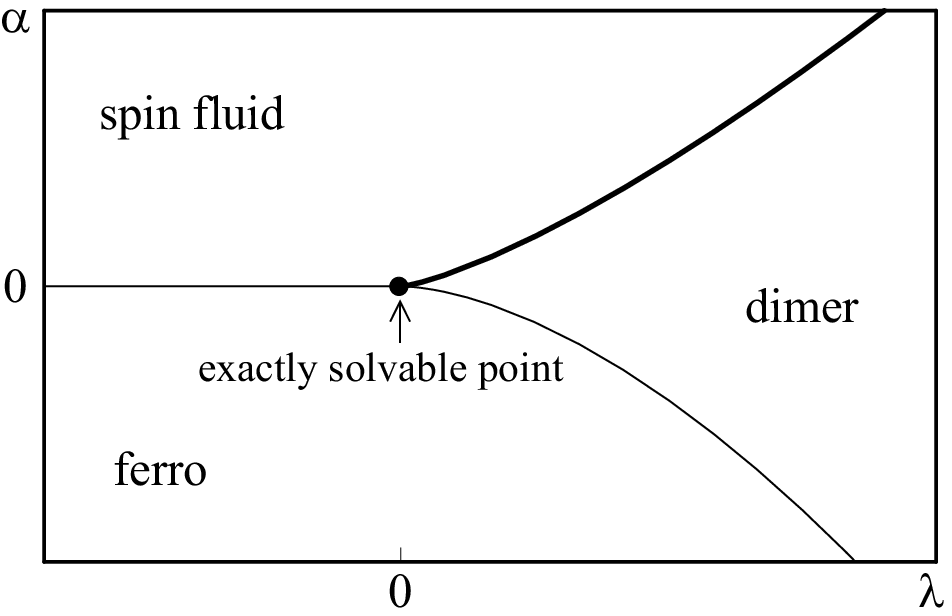}
\caption{Schematic phase diagram of the F-AF chain. 
The exactly solvable point is indicated by the dot at $\alpha = \lambda = 0$, and the phase boundary between the spin fluid and dimer phases is shown by the bold solid line.}
\label{fig:phase}
\end{center}
\end{figure}

The ground state near the exactly solvable point 
has also been examined. 
Several authors studied the issue by adding two deviation terms 
to the Hamiltonian at the exactly solvable 
point.\cite{tonegawa90,tonegawa92,somma01,dmitriev08} 
One is the anisotropy term with the anisotropy parameter $\alpha$, 
which displaces the exchange energy of the $z$-direction 
as $J_1(1-\alpha)$. 
The other is the NNN interaction term measured from the exactly solvable point; namely, the 
NNN coupling constant is 
the parameter $\lambda=J_2-|J_1|/4$.
The numerical studies show that the ground state for $\alpha>0$ and $\lambda=0$ belongs to the spin fluid (Tomonaga-Luttinger liquid) phase and that for $\alpha=0$ and $\lambda>0$ belongs to the dimer phase\cite{tonegawa90,tonegawa92,somma01}.
For $\alpha,\lambda \ll 1$, 
Dmitriev and Krivnov\cite{dmitriev08}
discussed a scaling property of the ground state energy.
They also argued about the phase boundary between the spin fluid and dimer phases and found the equation of the phase boundary as $\alpha_{\rm c}\simeq 13.9 \lambda_{\rm c}^{4/3}$ 
using the scaling property of the ground state energy in the dimer phase.
The sketch of the phase boundary
is shown in 
Fig.~\ref{fig:phase}.

In this paper, we investigate the ground state behavior of the anisotropic F-AF chain near the exactly solvable point. 
We consider the Hamiltonian of 
the F-AF chain at the exactly solvable point of $\alpha=\lambda=0$
as an unperturbed system and the other terms 
mentioned above as perturbations.
Using the recursion formulae of the exact solutions of the ground states, 
the ground state energy can be calculated within the first order term of the perturbation. 
From the result of the first order perturbation, 
we derive a scaling form of the ground state energy in the spin fluid phase.
This scaling form is confirmed numerically using the exact diagonalization. 
The detailed expression of 
the phase boundary between the spin fluid and dimer phases is determined from the scaling property with numerical data.

This paper is organized as follows.
In \S2, the Hamiltonian of the unperturbed system and the perturbation terms are explained.
In \S3, the exact solution of the ground state of the unperturbed system is given.
In \S4, within the first order perturbation theory, we discuss the phase boundary between the spin fluid and dimer phases.
In \S5, the detailed expression of the phase boundary is derived using the scaling property of the ground state energy.
Finally, \S6 devoted to summary and discussions.

\section{F-AF Chain with Perturbations}

We examine the effects of perturbations near the exact solutions for the F-AF chain at the exactly solvable point. 
The total Hamiltonian consists of the unperturbed Hamiltonian 
$\mathcal{H}_0$ and the perturbation terms $V_\lambda$ and 
$V_\alpha$: 
\begin{align}
 \mathcal{H}&=\mathcal{H}_0 + V_\lambda + V_\alpha,\label{eq:H}\\
 \mathcal{H}_0 &= \sum_{n} 
\left(-\boldsymbol{s}_n\boldsymbol{s}_{n+1} +\frac{1}{4} \boldsymbol{s}_n\boldsymbol{s}_{n+2}\right),
\label{eq:H_0}\\
 V_\lambda &=\lambda \sum_n\left( \boldsymbol{s}_n\boldsymbol{s}_{n+2} -\frac14 \right),
\label{eq:Vlambda}\\
 V_\alpha &= \alpha \sum_n s^z_n s^z_{n+1} , 
\label{eq:Valpha}
\end{align}
where $\boldsymbol{s}_n$ is the spin-1/2 operator at the $n$-th site,  
and the summations are taken over the total number $N$ of spin sites. 
We used the energy unit of $J_1=-1$, so that 
eqs.~(\ref{eq:H_0}) and (\ref{eq:Vlambda}) indicate $J_2 = \lambda + 1/4$. 

To treat the exact solutions for the unperturbed Hamiltonian 
$\mathcal{H}_0$ at the exactly solvable point, we write it as a sum of local Hamiltonians: 
$\mathcal{H}_0 = \sum_n h_n$, with 
\begin{align}
h_n = 
-\frac12 (\boldsymbol{s}_n\boldsymbol{s}_{n+1} + \boldsymbol{s}_{n+1}\boldsymbol{s}_{n+2} ) +\frac14 \boldsymbol{s}_n\boldsymbol{s}_{n+2} . 
\label{eq:h_n}
\end{align}
Here, the system size $N$ is assumed to be an even number. 
Owing to the form of the Hamiltonian, it is possible to use 
a special open boundary condition (OBC) with $n=1$ to $N-2$, 
shown in Fig.~\ref{fig:chain}, 
or a periodic boundary condition (PBC) with $n=1$ to $N$. 
The ground states under the OBC are degenerate with respect to 
all the magnitudes and $z$-components of the total spin. 
We have found the exact recursion formulae for all the degenerate ground states.\cite{hs08}
We choose the OBC, since we need all the degenerate ground states  to perform first-order perturbation calculations. 
The degeneracy of these ground states of $\mathcal{H}_0$ is 
resolved by the perturbation terms $V_\alpha$ and $V_\lambda$.

The perturbation term $V_\lambda$ changes the NNN interaction 
parameter $J_2$ from $1/4$ at the exactly solvable point to 
$\lambda + 1/4$. 
Accordingly, for $\lambda < 0$, 
the system shifts from the exactly solvable point 
into the ferromagnetic phase, if $\alpha = 0$. 
In contrast, for $\lambda > 0$, it shifts into the dimer phase.

The other perturbation term $V_\alpha$ represents an anisotropy of the NN interaction. 
If $V_\alpha$ is Ising-like ($\alpha<0$) and $\lambda=0$, 
the system shifts into the ferromagnetic phase where the total spin direction is along the $z$-axis. 
In contrast, if it is $XY$-like ($\alpha>0$), 
the system shifts into the spin fluid phase. 
In this case, the ground state is approximated by 
a linear combination of 
the degenerate ground states of $\mathcal{H}_0$ 
in the first-order perturbation. 
We concentrate on this interesting case of $\alpha>0$. 
Also, we consider the anisotropy only for the NN interaction 
and not for the NNN interaction for simplicity. 
The latter brings no drastic change as will be discussed. 

\section{Exact Solution for $\alpha=\lambda=0$}

We summarize the exact recursion formula 
for the unperturbed Hamiltonian $\mathcal{H}_0$ with the OBC, 
which describes the uniform and isotropic F-AF chain at the exactly solvable point. 
The general formula in the case with full bond alternations 
in the NN and NNN interactions is given 
in our precedent paper.~\cite{hs08} 

Let $|j,m \rangle_N$ be the ground state of the $N$-site 
chain for $S_{\rm tot}=j$ and $S_{\rm tot}^z=m$, 
where $S_{\rm tot}$ and $S_{\rm tot}^z$ are the magnitude and the $z$-component of the total spin $\boldsymbol{S}_{\rm tot} = \sum_{n=1}^{N} \boldsymbol{s}_n$, respectively.
The ground state $|j,m \rangle_{N+2}$ of the $(N+2)$-site chain 
is formed by incorporating spin states of two extra $1/2$ spins 
at the $(N+1)$-th and $(N+2)$-th sites. 
The triplet and singlet states for the extra spins are expressed as 
\begin{align}
&|{\rm t}_1\rangle = |\uparrow_{N+1} \uparrow_{N+2} \rangle ,\qquad 
 |{\rm t}_{-1}\rangle = |\downarrow_{N+1} \downarrow_{N+2} \rangle ,\nonumber\\
& |{\rm t}_0\rangle = \frac{1}{\sqrt2}(|\uparrow_{N+1} \downarrow_{N+2} \rangle 
	+ |\downarrow_{N+1} \uparrow_{N+2} \rangle),\nonumber\\
&
 |{\rm s}\rangle = \frac{1}{\sqrt2}(|\uparrow_{N+1} \downarrow_{N+2} \rangle 
	- |\downarrow_{N+1} \uparrow_{N+2} \rangle).
\end{align}
Then $|j,m \rangle_{N+2}$ is written as 
\begin{align}
|j,m\rangle_{N+2} =
& a_{N}(j) \sum_{\mu} 
{\rm C}(1,\mu)
| j+1,m-\mu \rangle_{N} \otimes |{\rm t}_\mu \rangle \nonumber\\
 + & b_{N}(j) \sum_{\mu} 
{\rm C}(0,\mu)
| j,m-\mu \rangle_{N} \otimes |{\rm t}_\mu \rangle\nonumber\\
 + & c_{N}(j) \sum_{\mu} 
{\rm C}(-1,\mu)
| j-1,m-\mu \rangle_{N} \otimes |{\rm t}_\mu \rangle\nonumber\\
 + & d_{N}(j) | j,m-\mu \rangle_{N} \otimes |{\rm s} \rangle 
\label{eq:N}
\end{align}
with the Clebsch-Gordan coefficients 
\begin{align}
{\rm C}(\nu,\mu) = \langle 1,\mu ; j+\nu,m-\mu | j,m \rangle 
\label{CGcoeff}
\end{align}
for $\mu = -1$, 0, and 1.  
The coefficients $\{a_{N}, b_{N}, c_{N}, d_{N}\}$ are 
independent of $m$ because of 
the rotational symmetry of $\mathcal{H}_0$. 
If an unreasonable state like $|(N/2)+1,m \rangle_{N}$ 
appears in the right-hand side of eq.~(\ref{eq:N}), 
we regard the coefficient for it as zero. 

We impose the condition that eq.~(\ref{eq:N}) is 
a common ground state of the local Hamiltonians 
$h_{N-1}$ and $h_{N}$. 
Then we obtain the recursion relations 
for $\{a_{N}, b_{N}, c_{N}, d_{N}\}$ 
with respect to the system size $N$ as 
\begin{align}
& \frac{a_{N}(j)}{d_{N}(j)} 
	= \frac{a_{N-2}(j)}{d_{N-2}(j+1)}, 
\label{eq:rec_a}\\
& \frac{b_{N}(j)}{d_{N}(j)} 
	= \frac{b_{N-2}(j)}{d_{N-2}(j)}, 
\label{eq:rec_b}\\
& \frac{c_{N}(j)}{d_{N}(j)} 
	= \frac{c_{N-2}(j)}{d_{N-2}(j-1)} .
\label{eq:rec_c}
\end{align}
Also, the following relations for the same $N$ are satisfied: 
\begin{align}
&\frac{b_N(j)}{d_N(j)}
= \sqrt{\frac{j-1}{j+1}} \, \frac{b_N(j-1)}{d_N(j-1)}
+4\sqrt{\frac{j}{j+1}}
\label{eq:b_N(j)}
\end{align}
and
\begin{align}
\frac{c_N(j)}{d_N(j)} 
  &  =       \left[\sqrt{j-1} b_N(j-1) + \sqrt{j} d_N(j-1)\right] \nonumber \\ 
  &  \times \left[\sqrt{j-1} b_N(j-1) + 3 \sqrt{j} d_N(j-1)\right] \nonumber \\
  &  \times \left[\sqrt{(2j+1)(2j-1)} \,a_N(j-1)d_N(j-1) \right]^{-1}.
\label{eq:c_N(j)} 
\end{align}
From the normalization condition of $|j,m\rangle_{N+2}$, we have
\begin{align}
a_{N}(j)^2+b_{N}(j)^2+c_{N}(j)^2+d_{N}(j)^2 = 1. 
\label{eq:normaliz} 
\end{align}

The recursion relations eqs.~(\ref{eq:rec_c})-(\ref{eq:normaliz}) 
determine the coefficients $\{a_{N}, b_{N}, c_{N}, d_{N}\}$ 
by starting from the initial values for $N=4$ 
except for coefficients with special values of $j$. 
Since the denominators of the right-hand sides of 
eq.~(\ref{eq:rec_a}) for $j=(N/2)-1$ and 
of eq.~(\ref{eq:rec_b}) for $j=N/2$ become zero, 
we separately evaluate $b_{N}(N/2)$ and $a_{N}((N/2)-1)$  
using eqs.~(\ref{eq:b_N(j)}) and (\ref{eq:c_N(j)}), respectively. 
The initial values of the coefficients and initial ground states are given in Appendix.

Using the above recursion relation with respect to the system size $N$, 
the matrix element $\langle j',m'|Q|j,m\rangle_n$ for an arbitrary $Q$ can be decomposed into the summation of the two-spin matrix element.
Thus, we can easily calculate an arbitrary correlation function for large $N$.

\section{First Order Perturbation}

The perturbation term $V_\alpha + V_\lambda$ 
in eq.~(\ref{eq:H}) resolves the degeneracy of the ground states 
of $\mathcal{H}_0$ with the OBC. 
We have considered the case of even $N$ for simplicity. 
We assume that the ground state of $\mathcal{H}$ 
exists in the subspace of $S_{\rm tot}^z=0$.  
Then, the zeroth order ground state $|\psi_0\rangle_N$ 
is written as a linear combination of the basis vectors of the subspace: 
\begin{align}
|\psi_0\rangle_N = \sum_{j=0}^{N/2} b_j |j,0\rangle_N.
\label{eq:phi0}
\end{align}
The coefficient vector 
$\boldsymbol{b} \equiv (b_0, b_1, \cdots , b_{N/2})^{\rm t}$ is 
the eigenvector for the smallest eigenvalue of 
the matrix $v$ whose $(j, j')$ element is defined as 
$v_{j,j'} = \langle j,0| V_\alpha + V_\lambda |j',0 \rangle_N$.
The first excited state $|\psi_1\rangle_N$ is also written 
in the same form as eq.~(\ref{eq:phi0}) but with the coefficient vector for the next smallest eigenvalue. 
The shift of the ground state energy for the first order perturbation is given by
\begin{align}
\delta E_{\alpha,\lambda }^{(1)}(N)=\langle \psi_0 | (V_\alpha +V_\lambda ) | \psi_0 \rangle_N , 
\end{align}
which is just the eigenvalue of $v$.

\begin{figure}[tbp]
\begin{center}
\includegraphics[width=80mm ]{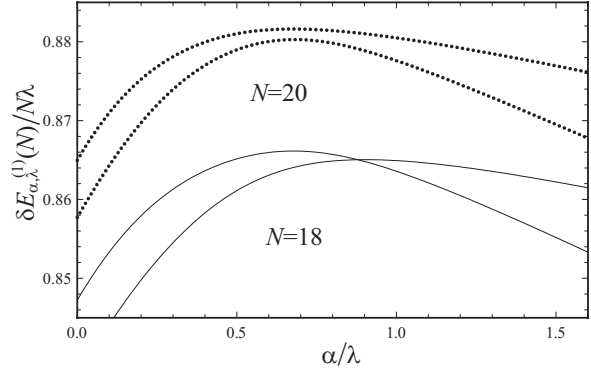}
\caption{Ground energy and first excitation energy for $N=18$ and $20$.}\label{fig:E_N18and20}
\end{center}
\end{figure}

The matrix element $v_{j,j'}$ is nonzero 
only for $j=j'$ or $j=j'\pm2$. 
Hence, we have $b_{2n}=0$ for all $n$, 
otherwise $b_{2n+1}=0$ for all $n$. 
Therefore, 
at least either $b_0$ or $b_{N/2}$ is zero if $N/2$ is odd, 
while both of them can be nonzero if $N/2$ is even.
For $\alpha=0$ and $\lambda> 0$ (dimer phase), eq.~(\ref{eq:phi0}) becomes $|\psi_0\rangle_N = |0,0\rangle_N$.
Hence, the state $|\psi_0\rangle_N$ for the dimer phase contains the state $|0,0\rangle_N$, namely $b_0\neq 0$. 
On the other hand, for $\alpha> 0$ and $\lambda = 0$ (spin fluid phase), eq.~(\ref{eq:phi0}) becomes $|\psi_0\rangle_N = |N/2,0\rangle_N$.
Hence, the state $|\psi_0\rangle_N$ for the spin fluid phase contains the ferromagnetic state $|N/2,0\rangle_N$, namely $b_{N/2}\neq 0$. 
Thus, for odd $N/2$, the eigenvector for the dimer phase 
and that for the spin fluid phase are orthogonal, 
so that the energy level crosses at the phase boundary. 
We illustrate the level crossing for $N=18$ in Fig.~\ref{fig:E_N18and20}.
This fact is observed not only in the first order perturbation but also in the numerical diagonalization of the total Hamiltonian with the OBC.
This is an effect of the OBC for a finite size system. 

\begin{figure}[tbp]
\begin{center}
\includegraphics[width=80mm ]{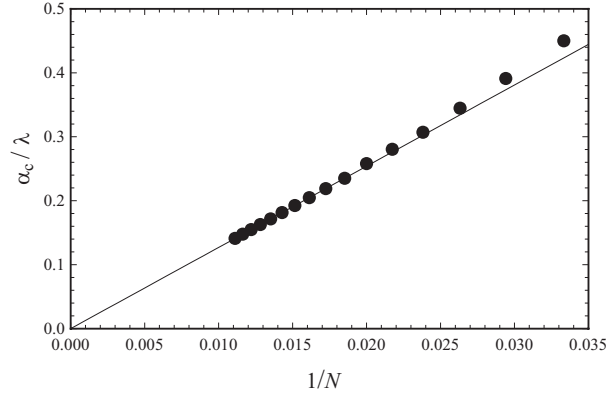}
\caption{Size dependence of the phase boundary calculated in first order perturbation. 
The line is obtained by fitting using data for $N=62\sim90$.}\label{fig:alpha_c_1st}
\end{center}
\end{figure}

The crossing points $\alpha_{\rm c}$ for $N\le 90$ are plotted in Fig.~\ref{fig:alpha_c_1st}. 
The solid line is obtained by fitting using data for $N=62\sim 90$.
We obtain $\alpha_{\rm c}\to 0$ in the thermodynamic limit $N \to \infty$. 
Therefore, $\alpha_{\rm c}$ can be expressed as a function of $\lambda $ in which the exponent of $\lambda$ is larger than 1.
This is consistent with the phase boundary given in ref.~\citen{dmitriev08} and the following section.

\section{Scaling Property and Phase Boundary}

We argue about the scaling property of the shift of the ground state energy for small $\alpha$ and $\lambda$ 
from the ground state energy $E_0$ of 
the unperturbed system $\mathcal{H}_0$. 
Before examining the general case correctly, 
we first treat the special case of $\lambda=0$ under the PBC, 
following the argument of Dmitriev and Krivnov\cite{dmitriev08}. 
Based on the perturbation argument, 
they introduced a scaling parameter $x$ as 
\begin{align}
x\sim \frac{\langle \psi_j| V_\alpha | \psi_k \rangle}{E_k-E_0}, 
\label{eq:x}
\end{align}
where $E_k$ is a typical low eigenenergy of $\mathcal{H}_0$ and $|\psi_k \rangle$ is the excited state belonging to $E_k$. 
In the most divergent contribution, 
the ground state energy and the corresponding eigenstate 
are respectively expanded in powers of $x$ as 
\begin{align}
\delta E_0(\alpha ) &= E_0(\alpha) -E_0 = \langle \psi_0|V_\alpha |\psi_0 \rangle \sum_i c_i x^i , 
\label{eq:expansion}
\\
|\psi_0(\alpha) \rangle &= \sum_k d_k x^{q_k}|\psi_k\rangle,
\label{eq:expansion_state}
\end{align}
where $c_i$'s and $d_k$'s are coefficients and $q_k$'s are non-negative integers. 
The 1-magnon excitation spectrum of $\mathcal{H}_0$ 
is written as\cite{harada90} 
\begin{align}
E_k - E_0 = (1-\cos k)-\frac14 (1-\cos 2k)\simeq \frac{k^4}{8}\sim \frac{1}{N^4}.
\end{align}
The numerator $\langle \psi_j| V_\alpha | \psi_k \rangle$ of $x$ 
is analyzed and found to be independent of $N$ 
for large $N$:\cite{dmitriev08}
\begin{align}
\langle \psi_j| V_\alpha | \psi_k \rangle \sim \alpha. 
\end{align}
Thus, the scaling parameter $x$ of the shift of the ground state energy is given by
\begin{align}
x = \alpha N^4.
\end{align}
Since the energy is 
an extensive variable, 
eq.~(\ref{eq:expansion}) can be written 
using the scaling function $f(x)$ as
\begin{align}
\delta E_0(\alpha ) = N\alpha^{5/4} f(x).
\label{eq:sc_alpha}
\end{align} 
This equation is confirmed by the 
numerical diagonalization with the PBC~\cite{dmitriev08}. 

\begin{figure}[tbp]
\begin{center}
\includegraphics[width=80mm ]{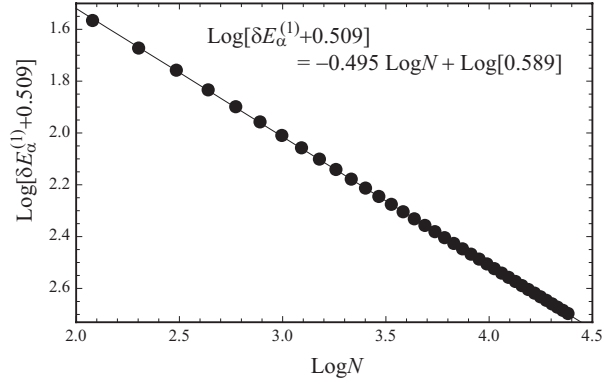}
\caption{Size dependence of $\delta E_\alpha^{(1)}(N)$.}\label{fig:dEalpha_N}
\end{center}
\end{figure}

\begin{figure}[tbp]
\begin{center}
\includegraphics[width=80mm]{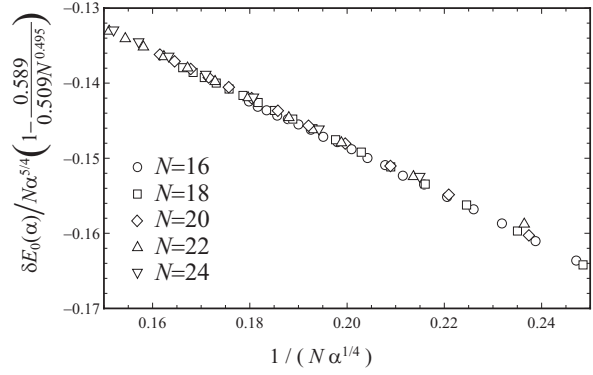}
\caption{Scaling function $f_{\rm OBC}(x)$ vs $1/x^{1/4}$.}
\label{fig:sc_alpha_open}
\end{center}
\end{figure}

To use the scaling form eq.~(\ref{eq:sc_alpha}) 
in our calculations, we need finite size correction 
in the case of the OBC. 
Using the exact recursion relations presented in the last section, 
we calculate $\delta E_0(\alpha)$ in the first order 
perturbation with respect to $\alpha$. 
The result is shown in Fig.~\ref{fig:dEalpha_N}. 
The energy shift is represented as a function of $N$ as 
\begin{align}
\delta E_\alpha^{(1)}(N)
& \equiv \lim_{\alpha\to 0}\frac{\delta E_0(\alpha)}{\alpha}  = \frac{\langle \psi_0| V_\alpha | \psi_0 \rangle_N }{\alpha} \nonumber\\
& = -0.509\left(1-\frac{0.589}{0.509\times N^{0.495}}\right).
\label{eq:dealpha_N}
\end{align} 
The $N$-dependent term in parentheses is the correction from the OBC, which is not negligible for $N\lesssim 24$. 
Therefore, the corrected energy shift has the following scaling form
\begin{align}
\frac{\delta E_0(\alpha )}{1-\displaystyle\frac{0.589}{0.509\times N^{0.495}}} = N \alpha^{5/4} f_\mathrm{OBC}(x). 
\end{align}
The scaling function obtained from the numerical diagonalization result for $N = 16 \sim 24$ with the OBC is shown in Fig.~\ref{fig:sc_alpha_open}. 


Now, we investigate the general case with both 
the perturbation terms, $V_\alpha$ and $V_\lambda$. 
Here, we assume that the parameter $\lambda$ takes a sufficiently small value so that the ground state belongs to the spin fluid phase. 
The scaling parameter $y$ caused by $V_\lambda$ is introduced as
\begin{align}
y\sim \frac{\langle \psi_j | V_\lambda | \psi_k \rangle}{E_k-E_0}. 
\end{align}
The $N$-dependence of the matrix element 
$\langle \psi_j | V_\lambda | \psi_k \rangle$ 
is written as 
\begin{align}
\langle \psi_j | V_\lambda | \psi_k \rangle\sim \lambda N^\theta 
\label{eq:matrix_lambda}
\end{align} 
with an exponent $\theta$, which will be determined later. 
Then the scaling parameter is represented as 
\begin{align}
y = \lambda N^{\theta+4}. 
\end{align} 
We do not adopt the formula $y=\lambda N^2$ that is used 
in ref.~\citen{dmitriev08}, since we find no adequate argument 
to justify it. 
Using the scaling function $f(x, y)$, 
the ground state energy is written as 
\begin{align}
\delta E_0(\alpha,\lambda) \equiv E_0(\alpha,\lambda ) -E_0 
= N \alpha^{5/4}f(x,y). 
\label{eq:fxy}
\end{align} 

\begin{figure}[tbp]
\begin{center}
\includegraphics[width=80mm ]{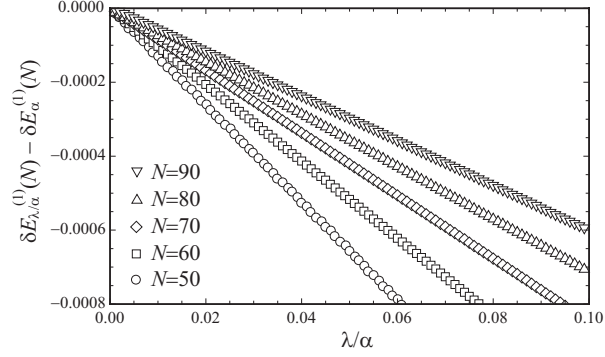}
\caption{$\lambda$ dependence of the ground state energy in the first order perturbation.}\label{fig:Vlambda}
\end{center}
\end{figure}

\begin{figure}[tbp]
\begin{center}
\includegraphics[width=80mm ]{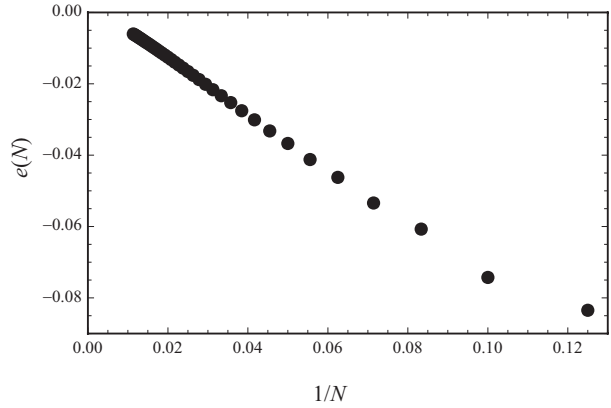}
\caption{Size dependence of the coefficient $e(N)$. }\label{fig:eN}
\end{center}
\end{figure}

To determine the exponent $\theta$, we consider the energy correction 
in the first order of $\lambda$. 
Using eqs.~(\ref{eq:expansion_state}) and 
(\ref{eq:matrix_lambda}), we have 
\begin{align}
E_0 (\alpha,\lambda) - E_0(\alpha,0) 
&= \langle \psi_0 (\alpha) |V_\lambda | \psi_0 (\alpha) \rangle 
\nonumber\\
&= \sum_{jk}d_j^* d_k x^{q_j+q_k} \langle \psi_j|V_\lambda| \psi_k \rangle 
\nonumber\\
&=\lambda N^\theta g(x),
\label{eq:dE0(a,l)}
\end{align} 
where $g(x)$ is a scaling function depending only on $x$.
We cannot evaluate 
$\langle \psi_0 (\alpha) |V_\lambda | \psi_0 (\alpha) \rangle$ directly 
in the first order perturbation with respect to $\alpha$, 
since we have no exact solutions for excited states 
to calculate the O($\alpha$) term of $| \psi_0 (\alpha) \rangle$.  
Therefore, we separately calculate $E_0 (\alpha,\lambda)$ and $ E_0(\alpha,0)$ in the first order of $\alpha$; 
the corrections are written as 
\begin{align}
E_0 (\alpha, 0) &= E_0 + \alpha \, \delta E_{\alpha}^{(1)}(N) + {\rm O}(\alpha^2) , \\
E_0 (\alpha,\lambda) &= E_0 + \alpha \, \delta E_{\lambda/\alpha}^{(1)}(N) + {\rm O}(\alpha^2) . 
\end{align} 
The former correction $\delta E_\alpha^{(1)} (N)$ has been given 
in eq.~(\ref{eq:dealpha_N}). 
We estimate the latter correction 
\begin{align}
& \delta E_{\lambda/\alpha}^{(1)}(N) \nonumber\\
& = \langle \psi_0 | \sum_n \left[ s_n^z s_{n+1}^z +\frac{\lambda}{\alpha}\left( \boldsymbol{s}_n\boldsymbol{s}_{n+2} - \frac14 \right) \right] | \psi_0 \rangle_N
\end{align} 
using the recursion relations in the last section.
The differences of the corrections for several values of $N$ 
are shown in Fig.~\ref{fig:Vlambda}. 
The result is represented in the form 
\begin{align}
\delta E_{\lambda/\alpha}^{(1)}(N)- \delta E_\alpha^{(1)} (N) = e(N)\frac{\lambda}{\alpha}.
\end{align}
The function $e(N)$ is linear for large $N$ as is shown in Fig.~\ref{fig:eN}.
Consequently, we have
\begin{align}
\langle \psi_0 (\alpha) |V_\lambda | \psi_0 (\alpha) \rangle \sim 
\lambda N^{-1},
\end{align}
and then the scaling parameter $y$ is given by 
\begin{align}
y=\frac{\lambda N^{-1}}{N^{-4}}=\lambda N^3.
\label{eq:y}
\end{align}


\begin{figure}[tbp]
\begin{center}
\includegraphics[width=80mm ]{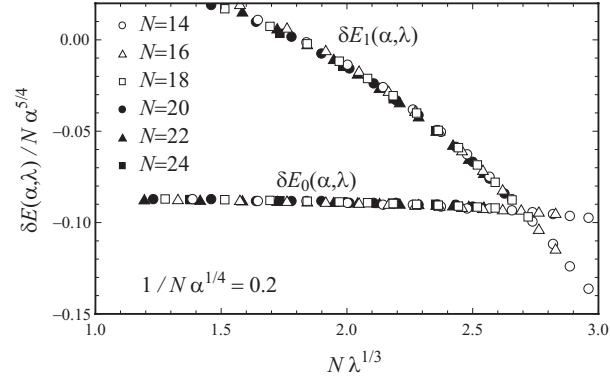}
\caption{
Scaling function $f(x,y)$ vs $y=N\lambda^{1/3}$ 
for a fixed value of 0.2 for $x = 1/(N\alpha^{1/4})$. 
The crossing point $(x_{\rm c}, y_{\rm c})$ corresponds to 
the phase boundary for $x_{\rm c}=0.2$.
}
\label{fig:scaling}
\end{center}
\end{figure}

We confirm the scaling form eq.~(\ref{eq:fxy}) 
by the numerical diagonalization under the PBC 
for finite size systems with $N=14\sim24$. 
The $y^{1/3}$ dependence of the scaling function $f(x,y)$ 
with a fixed $x$ for the ground state energy is shown in Fig.~\ref{fig:scaling}. 
Clearly, all the data lie well on a unique curve. 
We also show the same for the first excitation energy 
 $\delta E_1 (\alpha,\lambda)$. 
All the data also lie well on a curve. 
This means that the low excitation energy is 
also described by the same scaling parameters $x$ and $y$. 

\begin{figure}[tbp]
\begin{center}
\includegraphics[width=80mm ]{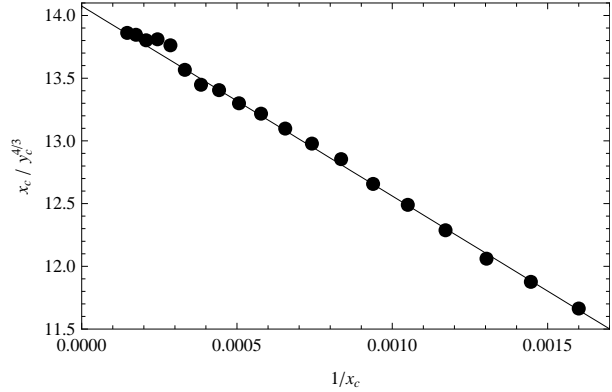}
\caption{
$x_{\rm c}$ dependence of ${x_{\rm c}}/{y_{\rm c}^{4/3}}$ for $N=22$.
}
\label{fig:pb}
\end{center}
\end{figure}

Extending the present scaling form for larger $\lambda$, 
we find a well-defined level crossing point $y_{\rm c}$ 
for each $x_{\rm c}$. 
This point corresponds to the phase boundary between the spin fluid and dimer phases. 
At this point, eliminating $N$ from $x_{\rm c}=\alpha_{\rm c} N^4$ and $y_{\rm c}=\lambda_{\rm c} N^3$, we have the following equation,
\begin{align}
\alpha_{\rm c} = \frac{x_{\rm c}}{y_{\rm c}^{4/3}} \lambda_{\rm c}^{4/3}.
\end{align}
Here, $y_{\rm c}$ is estimated by the numerical diagonalization for each fixed $x_{\rm c}$.
The $x_{\rm c}$ dependence of ${x_{\rm c}}/{y_{\rm c}^{4/3}}$ for $N=22$ is shown in Fig.~\ref{fig:pb}.
In the thermodynamic limit ($x=\alpha N^4 \to \infty$), we have ${x_{\rm c}}/{y_{\rm c}^{4/3}}\to 14 \pm 0.1$
and finally arrive at the following expression of the phase boundary:
\begin{align}
\label{eq:pb}
\alpha_{\rm c} \simeq 14 \lambda_{\rm c}^{4/3}.
\end{align}

\section{Summary and Discussion}

To summarize, we studied the phase boundary of 
the F-AF spin chain with an anisotropy and an extra NNN interaction, 
which are characterized by the parameters $\alpha$ and 
$\lambda = J_2 - |J_1|/4$, respectively. 
The analysis is perturbational with the exact solutions 
in the unperturbed case of $\alpha = \lambda =0$.
We investigated the scaling property of the ground state energy 
in the spin fluid phase.
We numerically confirmed the scaling form and 
arrived at the phase boundary equation 
between the spin fluid and dimer phases.

Here, we mention the following two points
in comparison with our results and the result of a previous study by Dmitriev and Krivnov\cite{dmitriev08}.
First, the scaling parameter $y$ obtained by Dmitriev and Krivnov ($y=\lambda N^2$) has a different expression from eq.~(\ref{eq:y}), because they adopted only the ferromagnetic ground state as a unperturbed state.
The scaling parameter of the ground state energy in the spin fluid state is correctly described by eq.~(\ref{eq:y}).
Second, our result of the phase boundary has the same form as the result they obtained.
We used a derivation based on the perturbation in the spin fluid phase, 
while they used the perturbation in the dimer phase.
Thus, eq.~(\ref{eq:pb}) is the definitive formula of the phase boundary between 
the spin fluid and dimer phases.

When the NNN interaction has the same anisotropy as that of the NN interaction with an $XY$-type, the similar scaling property discussed in \S5 can be obtained.
However, the scaling function plotted using the numerical diagonalzation for $N=16\sim20$ does not completely lie on one curve. 
If we use larger systems, the diagonalization data will lie on one curve.
The phase boundary estimated by the chain for $N=22$ is $\alpha_{\rm c} \simeq 15\lambda_{\rm c}^{4/3}$.

In the case of the Ising-type anisotropy of the NN interaction ($\alpha<0$), the first order phase transition between the dimer and ferromagnetic phases exists.
However, the detailed investigation using a finite size calculation is disturbed by the complicated size dependence\cite{tonegawa89} of the ground state energy in the dimer phase.
The meanfield approximation predicts that the phase transition takes place at $\alpha_{\rm c}\sim -\lambda_{\rm c}^{5/3}$\cite{dmitriev08}.
Within the first order perturbation used in \S4, we obtain the result that the linear term of $\alpha_{\rm c} $ with respect to $\lambda_{\rm c}$ becomes zero, which is consistent with the meanfield result.
In the case that the NNN interaction also has the same anisotropy as that of the NN interaction with an Ising-type, there appears an intermediate state between the dimer state and the fully polarized ferromagnetic state\cite{tonegawa01}.
Unfortunately, this intermediate phase does not appear by the first order perturbation.
The boundary between these phases can be roughly estimated by the numerical diagonalization, however, the complex size dependence disturbs the extrapolation of the system size $N$.
Detailed study in this regime remains in the future.

\section*{Acknowledgment}
This work is partly supported by a Fund for Project Research of Toyota Technological Institute.

\appendix
\section{Initial state of the recursion relation}
The smallest chain decomposed to the local Hamiltonian eq.~(\ref{eq:h_n}) is $N=4$ for even $N$.
Thus, the initial ground states for eq.~(\ref{eq:N}) are given as follows:
\begin{align}
&|2,2\rangle_4 
	= |\uparrow\uparrow\uparrow\uparrow\rangle, \nonumber\\
&|2,1\rangle_4 
	= \frac{(1,1,1,1) \cdot \boldsymbol{u}_1}{2}, \nonumber \\
& |1,1\rangle_4
	= \frac{(3 ,1, -1, -3) \cdot \boldsymbol{u}_1}
		{2\sqrt{5}}, \nonumber \\
& |2,0\rangle_4
	= \frac{(1,1,1,1,1,1) \cdot \boldsymbol{u}_0}{\sqrt6}, \nonumber \\
& |1,0\rangle_4
	= \frac{(2,1,0,0,-1,-2) \cdot \boldsymbol{u}_0}{\sqrt{10}}, \nonumber \\
& |0,0\rangle_4
	= \frac{(3,2,-3,-3,2,3 ) \cdot \boldsymbol{u}_0}
		{2\sqrt{9}}, \nonumber \\
& |2,-1\rangle_4
	= \frac{(1,1,1,1) \cdot \boldsymbol{u}_{-1}}{2}, \nonumber \\
& |1,-1\rangle_4
	= \frac{(3 ,1, -1, -3) \cdot \boldsymbol{u}_{-1}}
		{2\sqrt{5}}, \nonumber \\
& |2,-2\rangle_4 
 	= |\downarrow\downarrow\downarrow\downarrow\rangle, 
\label{eq:gs_4}
\end{align} 
where 
$\boldsymbol{u}_1 \equiv ( |\uparrow\uparrow\uparrow\downarrow\rangle 
, |\uparrow\uparrow\downarrow\uparrow\rangle
, |\uparrow\downarrow\uparrow\uparrow\rangle 
, |\downarrow\uparrow\uparrow\uparrow\rangle)^{\rm t}$, 
$\boldsymbol{u}_0\equiv(|\uparrow\uparrow\downarrow\downarrow\rangle 
, |\uparrow\downarrow\uparrow\downarrow\rangle
, |\downarrow\uparrow\uparrow\downarrow\rangle)
, |\uparrow\downarrow\downarrow\uparrow\rangle 
, |\downarrow\uparrow\downarrow\uparrow\rangle 
, |\downarrow\downarrow\uparrow\uparrow\rangle)^{\rm t}$ and 
$\boldsymbol{u}_{-1}\equiv( |\uparrow\downarrow\downarrow\downarrow\rangle
, |\downarrow\uparrow\downarrow\downarrow\rangle
, |\downarrow\downarrow\uparrow\downarrow\rangle
, |\downarrow\downarrow\downarrow\uparrow\rangle)^{\rm t}$.
The initial coefficients $\{a_4, b_4, c_4, d_4\}$ 
for the recursion relations eqs.~(\ref{eq:rec_a})-(\ref{eq:rec_c}) 
are given in Table~\ref{tab:N=4}.
\begin{table}[tbp]
\begin{center}
\caption{Initial coefficients $\{a_4, b_4, c_4, d_4\}$ 
for the recursion relations eqs.~(\ref{eq:rec_a})-(\ref{eq:rec_c}).}
\label{tab:N=4}
\begin{tabular}{c|cccc}
$j$ & 0 & 1 & 2 & 3\\
\hline
$a_4(j)$ 
	& $ \sqrt{15/17} $ 
	& $ \sqrt{30/77} $ 
	& 0 
	& 0 \\
$b_4(j)$ 
	& 0 
	& $ 2 \sqrt{10/77} $
	& $ 2 \sqrt{6/35} $
	& 0 \\
$c_4(j)$ 
	& 0 
	& $ \sqrt{2/77} $
	& $ \sqrt{2/7} $
	& 1 \\
$d_4(j)$ 
	& $ \sqrt{2/17} $ 
	& $ \sqrt{5/77} $
	& $ \sqrt{1/35} $
	& 0
\end{tabular}
\end{center}
\end{table}
Thus, the ground state $|j,m\rangle_N $ 
with arbitrary even $N$ is constructed recursively
by eqs. (\ref{eq:N}) and (\ref{eq:rec_a})-(\ref{eq:normaliz}). 
All the coefficients can be determined 
as positive values by a suitable choice of the sign of each ground state.
Thus, the coefficients $\{a_N,b_N,c_N,d_N\}$ are determined uniquely.
Therefore, the ground state in the sector of fixed $S_{\rm tot} = j$ and $S_{\rm tot}^z = m$ is nondegenerate. 
The total degeneracy of the ground states is 
$(N+2)^2/4$ for even $N$.


\end{document}